\documentclass[12pt,aps,a4paper, floats,nofootinbib,amssymb,superscriptaddress]{revtex4}
\usepackage[sort&compress]{natbib}
\usepackage{epsf,epsfig}
\usepackage{amsmath}
\usepackage{hyperref}
\usepackage{subfigure}
\usepackage{graphicx}
\usepackage{color}
\usepackage{mathrsfs}
\usepackage{enumitem}

\newcommand{\be}{\begin{equation}}
\newcommand{\ee}{\end{equation}}
\newcommand{\bea}{\begin{eqnarray}}
\newcommand{\eea}{\end{eqnarray}}

\newcommand{\ba}{\begin{array}}
\newcommand{\ea}{\end{array}}
\newcommand{\bi}{\begin{itemize}}
\newcommand{\ei}{\end{itemize}}

\newcommand{\bra}[1]{\langle #1 |}
\newcommand{\ket}[1]{| #1 \rangle}

\newcommand{\matelemgen}[3]{\bra{#2} #3 \ket{#1}}

\newcommand{\lag}{\mathcal{L}}

\newcommand{\myspace}{\vspace{0.2cm} \\}

\newcommand{\masspl}{M_{\text{pl}}}

\makeatletter
\begin{document}

\title{\vspace*{1.in}
 Unitarization of $R + \alpha R^2$ gravity
\vspace*{0.5cm}
}

%-----------------------------------
% Authors
%-----------------------------------
\author{I\~nigo  Asi\'ain}\email{iasiain@icc.ub.edu}\affiliation{Departament de F\'isica Qu\`antica i Astrof\'isica\,,
Institut de Ci\`encies del Cosmos (ICCUB), \\
Universitat de Barcelona, Mart\'i Franqu\`es 1, 08028 Barcelona, Spain}
\author{Antonio Dobado}\email{dobado@fis.ucm.es}\affiliation{Departamento de F\'isica Te\'orica, Instituto IPARCOS, \\
  Universidad Complutense, Plaza de las Ciencias 1, 28040 Madrid, Spain}
\author{Dom\`enec Espriu}\email{espriu@icc.ub.edu}
\affiliation{Departament de F\'isica Qu\`antica i Astrof\'isica\,,
Institut de Ci\`encies del Cosmos (ICCUB), \\
Universitat de Barcelona, Mart\'i Franqu\`es 1, 08028 Barcelona, Spain}

\thispagestyle{empty}

%-----------------------------------
% Abstract
%-----------------------------------
\begin{abstract}
  We make use of the improved K-matrix algorithm to obtain unitarized amplitudes in  $R+\alpha R^2$ gravity (the so-called
  Starobisnsky model, of cosmological relevance). The procedure is of some complexity because infrared divergences
  are present and need to be properly regulated. We focus on the behaviour of a bona fide scalar resonance, known to
  exist in this model, and compare it to an apparent resonance detected in previous studies, thus confirming
  that the latter seems to be an artifact due to the introduction of the infrared regulator. We analyze the existence of other
  dynamical resonances and dwell on the amplitudes made unitary by this procedure.
  \vspace*{0.7cm}
\noindent

\end{abstract}

\maketitle
\section{Introduction}
It is widely understood that non-renormalizable theories with a bad ultraviolet behaviour may exhibit dynamical resonances that,
when properly included, help mitigate their behaviour at high energies. Unitarization of the partial wave amplitudes is probably the preferred
method to unveil such resonances. In Ref.~\cite{Delgado:2022qnh,Delgado:2022uzu} the unitarization of Einstein-Hilbert gravity was carried out in the
framework of the Inverse Amplitude Method (IAM) and the one-loop level graviton-graviton scattering in pure gravity was unitarized.
Although the resulting amplitude was indeed unitary, we found no evidence of any genuine dynamical resonance, a fact that contradicted
the conclusions of a previous study regarding a singularity found in the second Riemann sheet in the complex $s$ plane
(termed as `graviball' by the authors in Ref.~\cite{Blas:2020dyg,Blas:2020och,Oller:2022ozd}). We concluded that such singularity -that actually lies quite
far away from the real $s$ axis anyway-  vanished as the infrared cut-off was removed.\myspace
In this work we want to extend this study to an extension of the Einstein-Hilbert Lagrangian; namely, one that includes in addtition
of the stardard term a $R^2$ piece. In fact, a gravity theory with a Lagrangian density of the form $R +\alpha R^2$ is known to possess a scalar
degree of freedom in addition to the familiar two helicities of the graviton. This scalar degree of freedom has been termed by some authors
the `scalaron'. One of our purposes is to show how the
unavoidable introduction of an infrared cut-off affects very differently both singularities (i.e. the graviball and the scalaron) reinforcing
the conclusions in Ref.~\cite{Delgado:2022qnh,Delgado:2022uzu}.\myspace
In addition we also want to explore possible physical consequences stemming from the unitarized partial waves and amplitudes, in particular
comparing with those of pure Einstein-Hilbert gravity. Although by no means a rigorous result,the emergence of a dynamical resonance would
probably signal the existence of new underlying microscopic degrees of freedom.\myspace
The approach followed here is consistent with assuming that gravity is an effective theory, with operators of increasing dimension that will
be expanded around flat minkowskian space. Specifically we will search for resonances that are commensurate with the Planck scale. In other words,
we do not consider the ultrahigh energy regime where graviton scattering could possibly be made unitary by appealing to black hole formation and
evaporation rather than relying on perturbative gravity. 
\section{The graviton}
From the perspective of quantum field theory, the graviton that mediates the gravitational interaction is a fundamental particle analogous
to the photon or the gauge bosons of the weak theory. It is a massless particle of spin $j=2$ that has
two degrees of freedom, corresponding to two possible physical helicity states of the graviton, $\lambda_{\pm} = \pm 2$. In order to
establish conventions and properly define our starting point we rederive the action for linearized gravity in a suitable gauge.
Our starting point is Einstein-Hilbert action:
\begin{equation}\label{eq_g_einsteinhilbert}
S_{EH} = \frac{1}{16\pi G} \int d^4x \sqrt{-g} R,  
\end{equation}  
where $R = g^{\mu\nu} R_{\mu\nu}$ is the Ricci scalar and $R_{\mu\nu}$ is the Ricci tensor. Both quantities are defined from the Rimeann tensor
and the Christoffel symbols. In our conventions
\begin{equation}\label{eq_g_RandChris}
\begin{aligned}
&R^{~\mu}_{\nu ~ \sigma\rho}=\partial_{\rho}\Gamma^{\mu}_{\nu\sigma}-\partial_{\sigma}\Gamma^{\mu}_{\nu\rho}+\Gamma^\mu_{\alpha\rho}\Gamma^\alpha_{\nu\sigma}-\Gamma^\mu_{\alpha\sigma}\Gamma^\alpha_{\nu\rho}\\
&\Gamma^{\rho}_{\mu\nu}=\frac{1}{2}g^{\rho\sigma}\left(\partial_\mu g_{\sigma\nu}+\partial_{\nu}g_{\sigma\mu}-\partial_{\sigma}g_{\mu\nu}\right),
\end{aligned}
\end{equation}
contracting one of the indices in the first line of Eq.~(\ref{eq_g_RandChris}) such that $ R_{\mu\nu}=R_{\nu}\, ^\sigma \, _{\sigma\nu}$.\myspace
We now expand  Eq.~(\ref{eq_g_einsteinhilbert}) around  Minkowski space-time,
\begin{equation}\label{eq_g_metricexpansion}
\begin{aligned}
&g_{\mu\nu}=\eta_{\mu\nu}+ h_{\mu\nu},\\
&g^{\mu\nu}=\eta^{\mu\nu}- h^{\mu\nu}+ h^{\mu\lambda}h_{\lambda}^\nu+\cdots
\end{aligned}
\end{equation}
with $\left|h^{\mu\nu}\right|\ll 1$ and the interpretation of  $h^{\mu\nu}$ as the graviton is immediate. \myspace
The expansion of the metric in Eq.~(\ref{eq_g_metricexpansion}) induces in turn an expansion of the Ricci tensor in terms
of $R^{(n)}_{\mu\nu}$, where $R^{(n)}_{\mu\nu}$ is of order $h^n$ (see Eq.~(\ref{eq_g_RandChris})), and the expanded Ricci scalar is
\begin{equation}
R=g^{\mu\nu}R_{\mu\nu}=\left(\eta^{\mu\nu}-h^{\mu\nu}+\cdots\right)\left(R^{(1)}_{\mu\nu}+R^{(2)}_{\mu\nu}+\cdots\right).
\end{equation}
With this the following action for the Einstein gravity in its linearized version is 
\begin{equation}\label{eq_g_SEH}
S_{EH}=-\frac{1}{64\pi G}\int d^4x\left[\partial_\mu h_{\alpha\beta}\partial^\mu h^{\alpha\beta}-\partial^\mu h \partial_\mu h+2\partial_\mu h^{\mu\nu}\partial_\nu h-2\partial_{\mu}h^{\mu\nu}\partial_\rho h^{\rho}_{\nu}\right],
\end{equation}
where we have defined $h\equiv h^\mu_\mu$, the trace of $h^{\mu\nu}$, and it has been used that $\sqrt{-g}\approx 1+\frac{1}{2}h+\mathcal{O}(h^2)$.\myspace
After choosing a reference frame where the expansion in Eq.~(\ref{eq_g_metricexpansion}) holds,
a residual symmetry of diff invariance remains
\begin{equation}\label{eq_g_residualgauge}
x^\mu\to x^{\prime\mu}=x^\mu+\xi^\mu(x)
\end{equation}
that in the linearized version of gravity implies invariance under
\begin{equation}\label{eq_g_gauge}
h_{\mu\nu} \to h_{\mu\nu} - \left( \partial_\mu\xi_\nu + \partial_\nu\xi_\mu \right). 
\end{equation}
The previous quadratic action can be written as 
\begin{equation}\label{eq_g_SPF}
S_2=\frac{1}{2}\int d^4x\left[-\partial_\rho h_{\mu\nu}\partial^\rho h^{\mu\nu}+2\partial_\rho h_{\mu\nu}\partial^\nu h^{\mu\rho}-2\partial_\nu h^{\mu\nu}\partial_\mu h+\partial^\mu h\partial_\mu h\right],
\end{equation}
 after integration by parts and the field rescaling:
$h_{\mu\nu}\to \left(32\pi G\right)^{-1/2}h_{\mu\nu}$. This rescaling means that now $h_{\mu\nu}$ should be thought of as a classical field with
canonical mass dimensions.\myspace
From now on the expansion of the metric on top of a Minkowski background will be 
\begin{equation}\label{eq_g_metricexpansionkappa}
\begin{aligned}
&g_{\mu\nu}=\eta_{\mu\nu}+\kappa h_{\mu\nu},\\
&g^{\mu\nu}=\eta^{\mu\nu}-\kappa h^{\mu\nu}+\kappa^2 h^{\mu\lambda}h_{\lambda}^\nu+\cdots
\end{aligned}
\end{equation}
where 
\begin{equation}
\kappa\equiv\left(32\pi G\right)^{1/2}\equiv\frac{\left(32\pi\right)^{1/2}}{M_{\text{pl}}}.
\end{equation}
\myspace
Of the ten degrees of freedom in the symmetric tensor $h_{\mu\nu}$, which represents the graviton field, only two physical degrees
of freedom remain after imposing the local symmetries. By applying the gauge invariance described in Eq.~(\ref{eq_g_gauge}), one can adopt the
Lorenz gauge\footnote{Also called the Hilbert gauge,  the harmonic gauge
or the De Donder gauge}
\begin{equation}\label{eq_g_lorentz}
\partial^\nu \tilde{h}_{\mu\nu} = 0 \longleftrightarrow \partial^\nu h_{\mu\nu}=\frac{1}{2}\partial_\mu h,
\end{equation}
where $\tilde{h}_{\mu\nu} = h_{\mu\nu} - \frac{1}{2} \eta_{\mu\nu} h$. These four equations reduce the degrees of freedom to six by choosing four
independent functions $\xi_\mu$ satisfying $\Box \xi_\mu = f_\mu (x)$ for any arbitrary $f_\mu (x)$ initial field configuration.
In this very same gauge, the equations of motion for the graviton in the linearized theory reduce to
\begin{equation}\label{eq_g_eomlinearnomatter}
\Box \tilde{h}_{\mu\nu}=0,
\end{equation}
which allows us to use four functions $\xi_\mu$ satisfying $\Box \xi^\mu = 0$ to further reduce the number of independent degrees of freedom
for the gaviton to two. These two physical degrees of freedom correspond to two independent helicity states, $\lambda_{\pm} = \pm 2$ that
we simply characterize as $\pm$.\myspace
Gauge fixing is required to find the propagator of the graviton field. The Lorentz gauge can be selected by introducing the following piece
\begin{equation}\label{eq_g_gaugefixing}
S_{gf}=-\int d^4x \left(\partial^\nu\tilde{h}_{\mu\nu}\right)^2,
\end{equation}
resulting in a propagator for the graviton $h_{\mu\nu}$ with momentum $k$
\begin{equation}
D_{\mu\nu\rho\sigma}(k)=\left(\eta_{\mu\rho}\eta_{\nu\sigma}+\eta_{\mu\sigma}\eta_{\nu\rho}-\eta_{\mu\nu}\eta_{\rho\sigma}\right)\left(\frac{-i}{k^2-i\epsilon}\right).
\end{equation}
After writing the gauge-fixing piece in Eq.~(\ref{eq_g_gaugefixing}) in terms of $h_{\mu\nu}$ and some integration by parts, one finds the action
\begin{equation}\label{eq_g_fullS2}
S=S_2+S_{gf}=\int d^4x \left[-\frac{1}{2}\partial_\rho h_{\mu\nu}\partial^\rho h^{\mu\nu}+\frac{1}{4}	\partial^\mu h \partial_\mu h\right].
\end{equation}
We will not consider the interaction with matter as we are only interested in pure gravity.\myspace
So much for the quadratic part of the action. Expansion around Minkowski space also introduces triple and quartic vertices
(and in fact vertices including any number of gravitons, but only the former will be required in the present study). Namely,
\begin{equation}\label{eq_g_Sfull}
  S=\int d^4x \left[-\frac{1}{2}\partial_\rho h_{\mu\nu}\partial^\rho h^{\mu\nu}+\frac{1}{4}	\partial^\mu h \partial_\mu h
    +\kappa \bar{S}_3+\kappa^2 \bar{S}_4+\cdots\right].
\end{equation}
where
\begin{equation}\label{eq_g_S3_and_S4}
\begin{aligned}
 &\bar{S}_3 = \frac{1}{2} \Bigg(
\frac{1}{4} h_{\mu}^{\;\mu} \, \partial_\nu h_{\alpha}^{\;\alpha} \, \partial^\nu h^{\;\beta}_{\beta}
- h^{\mu\nu} \, \partial_\mu h^{\alpha\beta} \, \partial_\nu h_{\alpha\beta} 
+ 2 h^{\mu\nu} \, \partial_\mu h^{\alpha\beta} \, \partial_\alpha h_{\nu\beta}
- \frac{1}{2} h^{\mu\nu} \, \partial_\alpha h_{\mu\nu} \, \partial^\alpha h^{\;\beta}_{\beta}\Bigg) \\ 
&\bar{S}_4 = \frac{1}{4} \Bigg(
-\frac{5}{16} h_\mu^{\;\;\mu} h_\nu^{\;\;\nu} \partial_\alpha h_\beta^{\;\;\beta} \partial^\alpha h^\tau_{\;\;\tau} + \frac{1}{2} h_\mu^{\;\;\mu} h^{\nu\alpha} \partial_\nu h_{\beta\tau} \partial_\alpha h^{\beta\tau} - h_\mu^{\;\;\mu} h^{\nu\alpha} \partial_\nu h^{\beta\tau} \partial_\beta h_{\alpha\tau}\\
& \qquad+ h_\mu^{\;\;\mu} h^{\nu\alpha} \partial_\beta h_{\nu\tau} \partial^\beta h_\alpha^{\;\;\tau} -\frac{1}{8} h_{\mu\nu} h^{\mu\nu} \partial_\alpha h_\beta^{\;\;\beta} \partial^\alpha h_\tau^{\;\;\tau}
+ h^{\mu\nu} \partial_\mu h_{\nu\alpha} \partial^\beta h^{\alpha\tau} h_{\beta\tau}\\
&\qquad + \frac{1}{4} h^{\mu\nu} \partial_\nu h_\alpha^{\;\;\alpha} h_{\nu\beta} \partial^\beta h_\tau^{\;\;\tau}
- 2 h^{\mu\nu} \partial_\mu h^{\alpha\beta} h_{\nu\alpha} \partial^\tau h_{\beta\tau} + h^{\mu\nu} \partial_\mu h_{\alpha\beta} h_{\nu\tau} \partial^\tau h^{\alpha\beta}\\
&\qquad - 2 h^{\mu\nu} \partial_\mu h_{\alpha\beta} h^{\alpha\tau} \partial_\tau h_\nu^{\;\;\beta} + h^{\mu\nu} h_{\nu\alpha} \partial_\beta h_{\mu\tau} \partial^\beta h^{\alpha\tau}
+ 2 h^{\mu\nu} \partial_\nu h_{\alpha\beta} h^{\alpha\beta} \partial^\tau h_{\mu\tau}
\Bigg).
\end{aligned}
  \end{equation}
Note that a mass term is conspicuously absent. It is actually forbidden by gauge invariance. The absence of a mass for the graviton
renders many processes infrared divergent. This is a combined effect of the graviton being a gauge boson, which exchanges in all
the three $s$, $t$ and $u$ channel and being strictly massless.\myspace
One may nevertheless be tempted to introduce
a mass term to regulate the omnipresent infrared divergences. A popular choice could be the Pauli-Fierz mass\cite{Fierz:1939ix}
\begin{equation}
\frac{m^2_g}{2}(h^{\mu\nu}h_{\mu\nu}-h^2).
\end{equation}
This choice allows five possible polarizations to propagate, but at least prevents the propagation of the Boulware-Deser
ghost (see Ref.~\cite{Alvarez-Gaume:2015rwa}),  typically present in massive spin two theories. However, it is known that
even in this case the $m_g\to 0$ limit is not smooth. This is the celebrated
van Dam-Veltmann-Zakharov discontinuity\cite{vanDam:1970vg,Zakharov:1970cc} which makes regulating the infrared with a mass
problematic.\myspace
Other methods have been used in dealing with the infrared singularities including dimensional regularization
and introducing a physical cut-off. In \cite{Donoghue:1999qh} it is shown that when including bremsstrahlung diagrams,
graviton cross section is finite and it depends on a scale that could be considered a physical cut-off separating IR and UV
regions.
Some effort with varying degree of success has been devoted to analyze whether these infrared singularities
exponentiate, see e.g. Ref.~\cite{Ware:2013zja}. They are certainly ubiquitous and
have to be dealt with one way or another. In what follows we will use a physical cut-off to deal with the IR singularities. 

\section{Effective Field Theories for gravitons}
The fact that the parameter $\kappa$ has dimensions of inverse energy, and that the action itself can be constructed as an expansion in
this parameter, $\kappa \sim 1/\masspl$, is a clear indication that the quantization of the theory of gravity leads to a non-renormalizable theory.\myspace
The resulting theory from this expansion can be rendered finite order by order in $\kappa$, requiring at any order only a finite, though not
necessarily small, new set of counterterms. From this perspective, we will be dealing anyway with an effective theory that describes the low-energy
effects of a more fundamental theory governed by an unknown dynamics, and whose leading-order term is the Einstein-Hilbert
action in Eq.~(\ref{eq_g_einsteinhilbert}) after the expansion of the metric.\myspace
From the previous considerations it becomes apparent that gravity shares some features with other effective theories based on a  derivative expansion,
such as the non-linear chiral Lagrangian describing the dynamics of massless Goldstone bosons in strong interactions, including both having massless
quanta and an analogous power counting. A discussion about the
possibility for the gravitons to be understood as Goldstones coming from a dynamical breaking of some UV theory can e.g. be
found in Refs.~\cite{Alfaro:2010ui, Alfaro:2012fs}, but several other attempts exist.\myspace
However, there is a notorious difference between gravity and the pion Lagrangian and it is that unlike the
chiral effective theory that is built upon a global symmetry,
the chiral symmetry, the Einstein-Hilbert theory has a local gauge symmetry that is responsible for the reduction of the number of degrees
of freedom in the low-energy spectrum and the presence of $t$ and $u$ exchange interactions.\myspace 
As in the case of the chiral Lagrangian, the next order in the expansion contributes to processes at $\mathcal{O}(p^4)$, and the set of
operators forming $\lag_4$ contains terms with four derivatives. In pure gravity---i.e., in the absence of matter and cosmological
constant ($\Lambda=0$)--- at the next order in the chiral counting, we can only construct the following action
\begin{equation}\label{eq_g_S4}
S_{\text{NLO}} = \int d^4x \sqrt{-g} \left[ \alpha_1 R^2 + \alpha_2 \left( R_{\mu\nu} \right)^2 + \alpha_3 \left(R_{\mu\nu\alpha\beta} \right)^2 \right].
\end{equation}
In fact, one can reduce these three operators to two, since they are related by topology through the Gauss-Bonnet term, which is a total derivative
(see for instance Ref.~\cite{Alvarez-Gaume:2015rwa}) 
\begin{equation}\label{eq_g_GBterm}
GB=R_{\mu\nu\rho\sigma}R^{\mu\nu\rho\sigma}-4R_{\mu\nu}R^{\mu\nu}+R^2.
\end{equation}
At the one-loop level, Einstein-Hilbert theory is free from ultraviolet divergences when the equations of motion ($R_{\mu\nu} = 0$) are applied.
This is as it should because the lowest order equation of motion force the two remaining operators in $\lag_4$ to vanish and one
would be left without any counterterms to eliminate a possible divergence, in accordance with the familiar effective field theory counting.\myspace
As a consequence, if one wants to go beyond Einstein-Hilbert gravity in the spirit of effective theories, the couplings $\alpha_i$ have to
be ``large'', meaning that the usual chiral counting is not applicable. This would also imply that the $\alpha_i \to 0$ limit is not necessarily
smooth (just like the $m_g \to 0$ limit in the Pauli-Fierz mass term is known not to be smooth either\cite{vanDam:1970vg,Zakharov:1970cc}).
Based on these considerations we abandon at this point the chiral Lagrangian inspired counting
and assume that the terms with two and four derivatives are equally important and on the same footing. This is in fact the logic behind
the so-called $f(R)$ theories. See Refs.~\cite{Sotiriou:2008rp, DeFelice:2010aj} for more information.\myspace
A particularly interesting case from the perspective of inflation, and the one chosen for our study, is given by  
\begin{equation}\label{eq_g_starobinski}
\begin{aligned}
&f(R)=R+\alpha R^2,\\
&S=\frac{1}{2\kappa^2}\int d^4x \sqrt{-g}\left(R+\alpha R^2\right),
\end{aligned}
\end{equation}  
known as the Starobinsky model, Ref.~\cite{STAROBINSKY198099}.\myspace
The equations of motion for $R$ with $f(R) = R + \alpha R^2$ are
\begin{equation}\label{eq_g_eoms}
R+6\alpha\Box R = 0,
\end{equation}  
which, for nonvanishing values of $\alpha$, immediately resembles the Klein-Gordon equation
\begin{equation}
\left(\Box+m^2\right) R=0,
\end{equation}
describing a free scalar field with a mass given by  
\begin{equation}\label{eq_g_scalaronmass}
 m^2 = \frac{1}{6\alpha}.
\end{equation}  
Indeed, by performing a conformal transformation  
\begin{equation}\label{eq_g_conformaltrans}
\tilde{g}_{\mu\nu} = \Omega^2 g_{\mu\nu}, \qquad \Omega^2 = \exp\left(\beta\varphi\right), \qquad \beta = \frac{1}{\masspl} \sqrt{\frac{16\pi}{3}},
\end{equation}  
to rewrite the action in Eq.~(\ref{eq_g_starobinski}) in the Einstein frame (instead of the Jordan frame, see Ref.~\cite{Maggiore:2007ulw}),
expressed in terms of quantities denoted with tildes, it is found that  
\begin{equation}
R = \Omega^2 \tilde{R} - 6 \Omega^{-1} \Box \Omega.
\end{equation}  
In this frame, the action becomes  
\begin{equation}\label{eq_g_actioneinsteinframe}
\begin{aligned}
&S = \int d^4x \sqrt{-\tilde{g}} \left( \frac{1}{2\kappa^2} \tilde{R} + \frac{1}{2} \partial^\mu \varphi \partial_\mu \varphi - V\left(\varphi\right) \right), \\
&V\left(\varphi\right) = \frac{3}{4\kappa^2} m^2 \left( \exp\left(-\beta\varphi\right) - 1 \right)^2,
\end{aligned}
\end{equation}  
where $m$ and $\beta$ are defined in Eqs.~(\ref{eq_g_scalaronmass}) and (\ref{eq_g_conformaltrans}), respectively.\myspace
Expanding the potential $V\left(\varphi\right)$ for $ \varphi / \masspl \ll 1 $,  
\begin{equation}\label{eq_g_potential}
V\left(\varphi\right) \approx \frac{1}{2}m^2\varphi^2 - 2\sqrt{\frac{\pi}{3}}\frac{m^2}{\masspl}\varphi^3 + \frac{14\pi}{9}\frac{m^2}{\masspl^2}\varphi^4 + \cdots,
\end{equation}  
we find that the action at next to leading order in Eq.~(\ref{eq_g_starobinski}), rewrites as the leading-order Einstein-Hilbert term plus a massive,
dynamical scalar field with self-interactions. This scalar field receives the name of \textit{scalaron}.\myspace
Recalling that $R\sim \partial\partial g\sim p^2$, $R^2$ corrections to Einstein-Hilbert will be comparable for energies such that  
\begin{equation}  
p^2\sim \alpha p^4 \Rightarrow s\sim \frac{1}{\alpha}=6m^2.  
\end{equation}
It is worth mentioning that when applying the Starobinsky model to cosmology, see Ref.~\cite{STAROBINSKY198099}, the scalaron mass of choice lies in the range:
$\sim 10^{-2}$ $M_{\rm pl}$ or less. Then
the coefficient $\alpha$, after pulling out a factor $\kappa/16\pi^2$, is definitely much larger than the natural order one would expect from
chiral counting (see e.g. \cite{Georgi:1984zwz}). This statement deserves some justification: once the theory is redefined in terms of the $\kappa$ parameter, $\kappa$ plays the role
of $v$, the vacuum expectation value in the electroweak theory. Then, the natural expansion parameter of the effective theory should be $\kappa^2/16\pi^2$, implying 
that the coefficient of the $p^4$ terms is $\sim 1/16\pi^2 \simeq 10^{-3}$ . In
the $R+\alpha R^2$ model, the coefficient of the $p^4$ term is $\alpha/2\kappa^2$ which is a number of order 1 if the mass is $10^{-2} M_{\rm pl}$, or even larger   
for lighter scalaron masses. 
\section{The spectrum of $R+\alpha R^2$}
We saw in the previous section that getting qualitatively different results from the ones of the Einstein-Hilbert requires somewhat large
values for $\alpha$. This brings us naturally into a non perturbative regime.\myspace
The scalaron, has a squared mass inversely proportional to the coupling constant $\alpha$ in the next-to-leading order effective theory of pure gravity,
as given by Eq.~(\ref{eq_g_scalaronmass}). The more significant the corrections to Einstein-Hilbert, the smaller the scalaron mass.
Conversely, the smaller the coupling constant $\alpha$ associated with the higher-order term, the larger the scalaron mass,
leading to greater decoupling from the pure gravity modes. The limit $\alpha \to 0$, where one might expect to recover as the mass tends to
infinity ($m \to \infty$), can be pathological due to the structure of the potential. In its expanded form, as shown in Eq.~(\ref{eq_g_potential}),
the potential includes interactions proportional to $m$ - a clear manifestation of non-decoupling. This places us in an scenario where unitarization
techniques can shed light on the high energy behaviour of the theory and  perhaps reveal some resonances.
We shall concentrate in the $J=0$ channel hereafter.\myspace
Previous works, such as those referenced in Refs.~\cite{Blas:2020dyg,Blas:2020och,Oller:2022ozd, Delgado:2022qnh, Oller:2024neq}, have explored the presence of scalar
resonances---and for higher values of $J$ in the case of Ref.~\cite{Delgado:2022qnh} in pure gravity and considering only the Einstein-Hilbert term.
We want to perform a similar study in the framework of the Starobinsky model.
\subsection{Relevant amplitudes}
\begin{figure}
\centering
\includegraphics[clip,width=6cm,height=3.5cm]{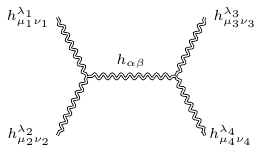}
\includegraphics[clip,width=6cm,height=3.5cm]{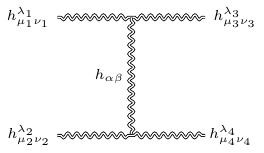}
\includegraphics[clip,width=6cm,height=3.5cm]{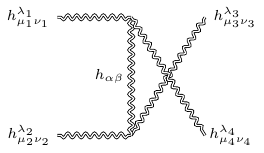}
\includegraphics[clip,width=6cm,height=3.5cm]{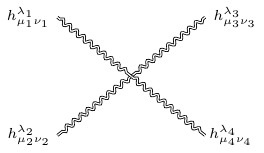}
\caption{\small{Diagrams contributing to the elastic scattering of gravitons in the helicity states $\lambda_i=\pm 2$.}}
\label{fig_g_diags_gggg}
\end{figure}

\begin{figure}
\centering
\includegraphics[clip,width=6.0cm,height=3.5cm]{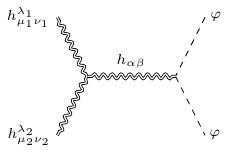}
\includegraphics[clip,width=6.0cm,height=3.5cm]{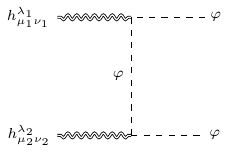}
\includegraphics[clip,width=6.0cm,height=3.5cm]{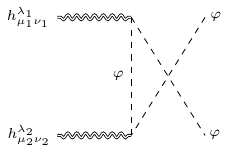}
\includegraphics[clip,width=6.0cm,height=3.5cm]{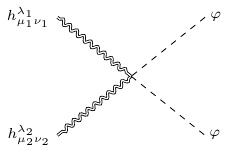}
\caption{\small{Diagrams contributing to the crossed process $hh\to \varphi\varphi$ where the gravitons are in the helicity states $\lambda_i=\pm 2$.}}
\label{fig_g_diags_ggss}
\end{figure}

\begin{figure}
\centering
\includegraphics[clip,width=6.0cm,height=3.5cm]{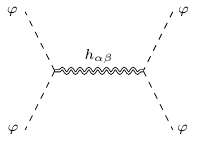}
\includegraphics[clip,width=6.0cm,height=3.5cm]{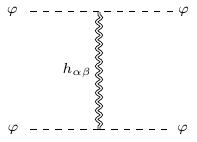}
\includegraphics[clip,width=6.0cm,height=3.5cm]{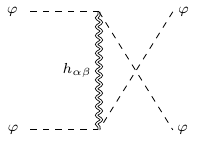}
\includegraphics[clip,width=6.0cm,height=3.5cm]{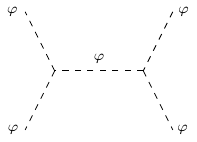}
\includegraphics[clip,width=6.0cm,height=3.5cm]{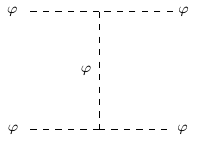}
\includegraphics[clip,width=6.0cm,height=3.5cm]{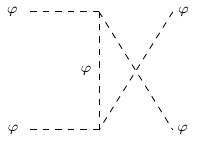}
\includegraphics[clip,width=6.0cm,height=3.5cm]{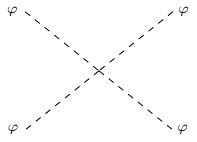}
\caption{\small{Diagrams contributing to the elastic scattering of scalarons.}}
\label{fig_g_diags_ssss}
\end{figure}

To study the scalar projection of $2 \to 2$ processes, three relevant amplitudes come into play: $h^{\lambda_1}_{\mu_1\nu_1} h^{\lambda_2}_{\mu_2\nu_2} \to h^{\lambda_3}_{\mu_3\nu_3} h^{\lambda_4}_{\mu_4\nu_4}$, $h^{\lambda_1}_{\mu_1\nu_1} h^{\lambda_2}_{\mu_2\nu_2} \to \varphi \varphi$, and $\varphi \varphi \to \varphi \varphi$. These are
schematically represented as $\mathcal{A}^{hhhh}$, $\mathcal{A}^{hh\varphi\varphi}$, and $\mathcal{A}^{\varphi\varphi\varphi\varphi}$, respectively.\myspace
In general, we define the momenta and helicities---for the gravitons---of the incoming particles as $p_1$, $\lambda_1$, $p_2$, $\lambda_2$, and the
outgoing particles as $p_3$, $\lambda_3$, $p_4$, $\lambda_4$. The helicity amplitudes are then expressed as 
\begin{equation}
\begin{aligned}
&\mathcal{A}^{hhhh}_{\lambda_1, \lambda_2, \lambda_3, \lambda_4} = \matelemgen{p_1, \lambda_1; p_2, \lambda_2}{p_3, \lambda_3; p_4, \lambda_4}{\mathcal{A}^{hhhh}}, \\
&\mathcal{A}^{hh\varphi\varphi}_{\lambda_1, \lambda_2} = \matelemgen{p_1, \lambda_1; p_2, \lambda_2}{p_3; p_4}{\mathcal{A}^{hh\varphi\varphi}}, \\
&\mathcal{A}^{\varphi\varphi\varphi\varphi} = \matelemgen{p_1; p_2}{p_3; p_4}{\mathcal{A}^{\varphi\varphi\varphi\varphi}},
\end{aligned}
\end{equation}
where $\lambda_i$ takes values $\pm 2$, denoted as $\pm$. The usual Mandelstam relations are defined as $s = (p_1 + p_2)^2$, $t = (p_1 - p_3)^2$, and $u = (p_1 - p_4)^2$.\myspace
Now, we present the explicit amplitudes for the three processes under study in this section. All of them are computed at tree level using the
FeynGrav program (Ref.~\cite{Latosh:2023zsi}). For each helicity combination, the diagrams involved in the elastic graviton scattering process include
the $s$-, $t$-, and $u$-channel diagrams with a graviton exchange and a four-graviton contact diagram and are represented in Fig.~\ref{fig_g_diags_gggg}.
The explicit sixteen decomposed helicity amplitudes $\mathcal{A}^{hhhh}_{\lambda_1, \lambda_2, \lambda_3, \lambda_4}$ are gathered in Table~\ref{tab_g_elasticgraviton}. 
\begin{table}[h!]
\centering
\renewcommand{\arraystretch}{1.5}
\begin{tabular}{|c|c|c|c|c|}
\hline
  & \(\ket{+; +}\) & \(\ket{+; -}\) & \(\ket{-; +}\) & \(\ket{-; -}\) \\ \hline
\(\bra{+; +}\) & \(\frac{\kappa^2}{4} \frac{s^3}{tu}\) & \(0\) & \(0\) & \(0\) \\ \hline
\(\bra{+; -}\) & \(0\) & \(\frac{\kappa^2}{4} \frac{u^3}{st}\) & \(\frac{\kappa^2}{4} \frac{t^3}{su}\) & \(0\) \\ \hline
\(\bra{-; +}\) & \(0\) & \(\frac{\kappa^2}{4} \frac{t^3}{su}\) & \(\frac{\kappa^2}{4} \frac{u^3}{st}\) & \(0\) \\ \hline
\(\bra{-; -}\) & \(0\) & \(0\) & \(0\) & \(\frac{\kappa^2}{4} \frac{s^3}{tu}\) \\ \hline
\end{tabular}
\caption{Table summarizing the sixteen helicity combinations for elastic graviton scattering.}
\label{tab_g_elasticgraviton}
\end{table}\myspace
Parity ($T_{\lambda_1,\lambda_2,\lambda_3,\lambda_4}=T_{-\lambda_1,-\lambda_2,-\lambda_3,-\lambda_4}$) and time-reversal invariance ($T_{\lambda_1,\lambda_2,\lambda_3,\lambda_4}=T_{\lambda_3,\lambda_4,\lambda_1,\lambda_2}$), corresponding to $PT$ symmetry, are explicitly shown in Table~\ref{tab_g_elasticgraviton}, whose entries are fully symmetric.\myspace
The crossed channel, i.e., $h^{\lambda_1}_{\mu_1\nu_1}h^{\lambda_2}_{\mu_2\nu_2}\to \varphi \varphi$, has contributions from four diagrams that are depicted
in Fig.~\ref{fig_g_diags_ggss}.  There is a $s$-channel graviton
exchange, followed by two diagrams in the $t$- and $u$-channels with a scalar exchange. Finally, there is a contact vertex involving two gravitons
and two scalarons. Four helicity combinations are possible, summarized as  
{\small
\begin{equation}
\mathcal{A}^{hh\varphi\varphi}=-\frac{\kappa^2}{64s^2}\left(2 s^3 (h_1 h_2 + 1) + 2 s (h_1 h_2 + 1) (t - u)^2 - \frac{s \left(s^2 + 2 s (t + u) + (t - u)^2\right)^2}{\left(m^2 - t\right) \left(m^2 - u\right)}\right),
\end{equation}}
where $h_{1,2}$ represents the helicity with a normalization such that $h_{1,2}=\pm 1$.\myspace
For elastic scattering of scalarons, seven types of diagrams are involved. Four of these, namely the $s$-, $t$-, $u$-channel, and contact
diagrams with scalar exchange, are identical to the equivalent Higgs processes described by anomalous triple and quartic selfinteractions $\lambda_3 v h^3$ and $\lambda_4 h^4$, where $v$ is the vev of the Higgs. These Higgs amplitudes are easily found in the literature (see for instance Eq. (17) of Ref.~\cite{Asiain:2021lch}) and the corresponding scalaron amplitudes are obtained straightforwardly through the following replacements:
\begin{equation}
  \lambda_3\to -2 \sqrt{\frac{\pi}{3}} \frac{m^2}{\masspl v}, \qquad \lambda_4\to\frac{56 \pi}{9} \frac{m^2}{\masspl^2}
\end{equation}
for the triple and quartic couplings, respectively.\myspace
The remaining three diagrams correspond to the $s$-, $t$-, and $u$-channels with graviton exchange. All of them together are shown in
Fig.~\ref{fig_g_diags_ssss}.
The total amplitude for the elastic scattering of scalarons is
{\small
\begin{equation}
\begin{aligned}
&\mathcal{A}^{\varphi\varphi\varphi\varphi}=\frac{1}{12} \kappa ^2 \left(18 m^4 \left(\frac{1}{m^2-s}+\frac{1}{m^2-t}+\frac{1}{m^2-u}\right)-14 m^2-\right.\\
&\left.\frac{3}{8stu}\left(-s^2 \left(6 t^2+7 t u+6 u^2\right)+s^3 (t+u)+s (t+u) \left(t^2-8 t u+u^2\right)+t u \left(t^2-6 t u+u^2\right)\right)\right).
\end{aligned}
\end{equation}}
With these amplitudes, we can construct the associated partial waves using
\begin{equation}\label{eq_generalpartialwaves}
t_J^{\lambda_1\lambda_2\lambda_3\lambda_4}(s)=\frac{1}{32K\pi}\int_{-1}^{1}d\left(\cos\theta\right) W_J^{\lambda,\lambda^\prime}\left(\cos\theta\right)\mathcal{A}\left(s,\cos\theta\right),
\end{equation}
where $\lambda = \lambda_1 - \lambda_2$ and $\lambda^\prime = \lambda_3 - \lambda_4$. As mentioned earlier, we will focus on the $++++$ amplitudes,
so $\lambda = \lambda^\prime = 0$, and the Wigner functions $W_J^{00}$ reduce to Legendre polynomials, which for the $J = 0$ projection
is $P_0\left(\cos\theta\right) = 1$.
For the elastic graviton-graviton case, the partial waves are not well-defined even at tree level, as can be directly seen in
Table~\ref{tab_g_elasticgraviton}, since they diverge for $t \to 0$ ($\cos\theta \to \pm 1$), requiring an infrared regularization.\myspace
Thus, the integral in Eq.~(\ref{eq_generalpartialwaves}) is redefined over $\cos\theta \in \left[-1 + \eta, 1 - \eta\right]$, with $0<\eta < 1$,
and the physical region where the partial wave is defined is $s > \mu^2$. These two free parameters are necessarily related because crossing
symmetry requires $t < -\mu^2$, as noted by the authors in Ref.~\cite{Delgado:2022qnh,Delgado:2022uzu}, taking into account that for massless external
states $t=-s/2(1-\cos\theta)$ so
\begin{equation}
\int_{-1 + \eta}^{1 - \eta} d\cos\theta = \frac{2}{s} \int_{-s + \mu^2}^{-\mu^2} dt,
\end{equation}
from where it follows that
\begin{equation}
\mu^2 = \eta \frac{s}{2}.
\end{equation}
This procedure guarantees perturbative unitarity (see e.g. Ref.~\cite{Delgado:2022qnh,Delgado:2022uzu}).\myspace
The result of this integration for the three processes gives
\begin{equation}
t_0^{hhhh}=\frac{s}{2\masspl^2}\log\left(\frac{2}{\eta}\right)
\end{equation}
\begin{equation}
t_0^{hh\varphi\varphi}=-\frac{s+8 m^2}{24 \masspl^2}+\frac{m^4}{2\sigma\masspl^2 s}\log\left(\frac{1+\sigma}{1-\sigma}\right)
\end{equation}
{\small
\begin{equation}
\begin{aligned}
t_0^{\varphi\varphi\varphi\varphi}=& \frac{1}{24\masspl^2 s \left(s-m^2\right)\left(s-4m^2\right)}\left[s^4\left(12 \log \left(\frac{2}{\eta }\right)-11\right)+m^2 s^3 \left(55-60 \log \left(\frac{2}{\eta }\right)\right)\right.\\
&\left.-4 m^4 s^2 \left(19-18 \log \left(\frac{2}{\eta }\right)-18 \log \left(\frac{s-3 m^2}{m^2}\right)\right)-4m^6s\left(-31+6\log\left(\frac{2}{\eta}\right)\right.\right.\\
&\left.\left. -18\log\left(\frac{s-3m^2}{m^2}\right)\right)+16m^8\right],
\end{aligned}
\end{equation}}
where $\sigma$ is the two-body phase space $\sigma=\sqrt{1-\frac{4m^2}{s}}$.\myspace
All of them are gathered in a $t_0$ scalar wave matrix defined as
\begin{equation}\label{eq_g_t0matrix}
t_0=\begin{pmatrix}
t_0^{hhhh} & t_0^{hh\varphi\varphi}\\
t_0^{hh\varphi\varphi} & t_0^{\varphi\varphi\varphi\varphi}
\end{pmatrix},
\end{equation}
that is suitable for the unitarization of the processes in the context of the coupled-channel formalism.
According to this unitarization mechanism, using matrix notation, we refer to the elastic scattering of gravitons ($hhhh$), as the $11$ channel
(corresponding to the $11$ entry Eq.~(\ref{eq_g_t0matrix}), and to the elastic scattering of scalarons ($\varphi\varphi\varphi\varphi$), as
the $22$ channel. This notation will be used for both unitarized and non-unitarized processes.\myspace
It is interesting to note that in the $m=0$ limit (i.e. formally taking $\alpha\to\infty$) the $t_0^{hhhh}$ partial wave coincides
with $t_0^{\varphi\varphi\varphi\varphi}$ and fully 
dominated by the IR log. On the other hand, $t_0^{hh\varphi\varphi}$ is free of IR singularities
\section{Unitarization and Riemann sheets}
\begin{figure}
\centering
\includegraphics[width=0.47\textwidth]{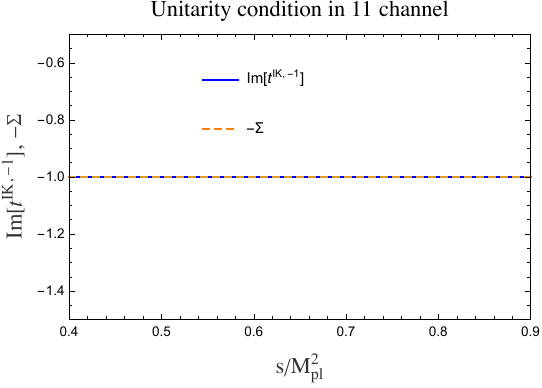}
\includegraphics[width=0.47\textwidth]{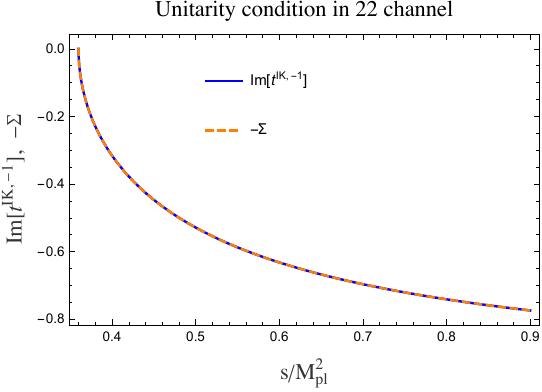} 
\caption{\small{Both sides of the full unitarity condition in Eq.~(\ref{eq_g_unitcondition}) for the values $m=0.3\masspl$ and $\mu=10^{-10}\masspl$.
    Both lines coincide meaning that the scalar waves fulfill unitarity as expected}}
\label{fig_g_unitconditions}
\end{figure}

The unitarization method chosen here is the K-matrix in its improved version, the IK-matrix (IK), is described in e.g. Ref.~\cite{Delgado:2015kxa}
(see also Ref.~\cite{Salas-Bernardez:2026gda} for a recent review and Ref.~\cite{Oller:2020guq} for apllications to hadronic physics). This unitarization method admits amplitudes calculated solely at tree level, adding by hand a physical---also unitarity---cut for the relevant continuations. In contrast, the IAM requires
the calculation of the amplitude at next-to-leading order for all three amplitudes. These amplitudes are not currently available and are exceedingly
laborious to compute. To the best of our knowledge, no literature exists to date that includes these amplitudes with a selfinteracting scalar contribution
at the one-loop level. Only the contribution from pure gravity is available, see Refs.~\cite{Dunbar:1994bn,Dunbar:1995ed,Delgado:2022qnh,Delgado:2022uzu}.\myspace
Therefore, the IK method turns out to be the most appropriate for a first approximation to the study of possible dynamical resonances in the $R+\alpha R^ 2$  theory.\myspace
The presence of both massive particles, such as the scalaron, and massless particles, such as the graviton, in the asymptotic states of the processes
for the unitarization of the $J=0$ channel leads to an analytical structure of the amplitudes with different cuts on the real axis. Through these cuts,
the different partial waves can be analytically continued to access the different Riemann sheets. In general, with $n$ coupled channels, $2^n$ Riemann
sheets open up.\myspace
In our case, we have two unitarity cuts: the first at the origin, described by the usual logarithm $L(s)$, and the second at $\text{Re}\,s = 4m^2$,
where $m$ is the mass of the scalaron as defined in Eq.~(\ref{eq_g_scalaronmass}), described by the Chew-Mandelstam function $J(s)$. All the possible
combinations of continous crossings through these two physical cuts result in four Riemann sheets. The explicit expressions for $L(s)$ and $J(s)$ are:
\begin{equation}\label{eq_L_and_J}
\begin{aligned}
&L(s)=\log(-s/\Lambda^2)\\
&J(s)=\sigma\log\left(\frac{\sigma+1}{\sigma-1}\right),
\end{aligned}
\end{equation}
and the corresponding continuation from the first to the second Riemann sheet (2RS):
\begin{equation}\label{eq_g_continuedL_and_J}
\begin{aligned}
&L^{\text{II}}(s)=\log\,|s|+i\left( \text{arg}\, s - \pi\right)\\
&J^{\text{II}}(s)=\sigma\log\,\left|\frac{\sigma+1}{1-\sigma}\right|+ i \sigma\left(\text{arg}\,\frac{\sigma+1}{1-\sigma}-\pi \right),
\end{aligned}
\end{equation}
where the superindex $\text{II}$ indicates that the function is taken in the second Riemann sheet.\myspace
These choices guarantee that, out of the possible $2^n$ sheets, one seeks for resonances in the so-called adjacent sheet---the one continously connected
to the physical region. In the present case with $n=2$, four different Riemann sheets open up, but the sheet we refer to as second Riemann
sheet---where physical resonances are located---depends on $\text{Re}\,s$. For $\text{Re}\,s<4m^2$, no continuation of $J(s)$ is required, since
the second cut is not reachable, and the 2RS is defined by continuing only $L(s)$. Conversely, for $\text{Re}\,s>4m^2$, the adjacent sheet is achieved
when both functions $L(s)$ and $J(s)$ are continued according to Eq.~(\ref{eq_g_continuedL_and_J}).\myspace
In summary, the IK-matrix method in its coupled-channel version provides the following prescription:
\begin{equation}\label{eq_t0}
t_{0}^{\text{IK}}=\left(1+t_{0}^{(2)}\cdot G (s)\right)^{-1} t_{0}^{(2)}
\end{equation}
where $1$ represents the identity matrix and $t_{0}^{(2)}$ is the scalar wave matrix built solely with the leading-order Lagrangian---recall the
tree-level calculation in this work---and is defined in Eq.~(\ref{eq_g_t0matrix}). $G(s)$ is a matrix containing all the unitarity cuts for the
different processes. In  our case, 
\begin{equation}\label{eq_un_Gmatrix}
G(s) = \frac{1}{\pi} \begin{pmatrix}
L(s) & 0 \\
0 & g(s)
\end{pmatrix}, \quad g(s) = \log\frac{m^2}{\Lambda^2} + J(s),
\end{equation}
with $L(s)$ and $J(s)$ defined in Eq.~(\ref{eq_L_and_J}).\myspace 
A straightforward calculation is needed to show that Eq.~(\ref{eq_t0}) with the function $G(s)$ in Eq.~(\ref{eq_un_Gmatrix}) indeed fulfills
all requirements from unitarity,
\begin{equation}\label{eq_g_unitcondition}
\text{Im}\,t_{0}^{\text{IK},-1}=\left(t_0^{(2)}\right)^{-1}\left(1+t_0^{(2)}\cdot G(s)\right)=-\Sigma,\quad \Sigma = \begin{pmatrix}
1 \quad & 0 \, \\
0 \quad & \sigma
\end{pmatrix}
\end{equation}
since $\text{Im}\, g(s)=-\pi\sigma$ in the physical region.\myspace
The matrix $G(s)$ contains the only functions that are to be continued to the second Riemann sheet for searching poles using Eq.~(\ref{eq_g_continuedL_and_J}),
as tree-level amplitudes lack an imaginary part. Thus, together with the previous discussion, the 2RS is uniquely defined everywhere:
\begin{equation}
G^{\text{II}}\left(s<4m^2\right)= \frac{1}{\pi} \begin{pmatrix}
L^{\text{II}}(s) & 0 \\
0 & g(s)
\end{pmatrix}, \quad G^{\text{II}}\left(s>4m^2\right)= \frac{1}{\pi} \begin{pmatrix}
L^{\text{II}}(s) & 0 \\
0 & g^{\text{II}}(s)
\end{pmatrix},
\end{equation}
where $g^{\text{II}}=\log\frac{m^2}{\Lambda^2} + J^{\text{II}}(s)$.\myspace
The single-channel approximation---i.e. neglecting the mixing with scalaron contributions---can be obtained directly from the expression
in Eq.~(\ref{eq_t0}) by replacing all matrices with their corresponding algebraic functions describing graviton-graviton scattering, yielding
\begin{equation}\label{eq_t0_sc}
t_0^{\text{IK-SC}}=\frac{t^{(2)}_0}{1+t_0^{(2)}L(s)}.
\end{equation}
Apart from the necessary checks to make contact with previous studies, in this work we adhere to the coupled-channel formalism in order to provide
an accurate description of scalar resonances in $R+ \alpha R^2$.\myspace
To conclude this section we demonstrate the unitarity of our partial waves after aplying the IK method in Fig.~\ref{fig_g_unitconditions}, where
we show both sides of Eq.~(\ref{eq_g_unitcondition}) for the scalaron mass $m=0.3 \masspl$ and $\mu=10^{-10}\masspl$. The lhs is displayed as a blue
solid line and the opposite-sign phase-space matrix in the rhs in dashed orange. Both lines coincide exactly revealing that full-unitarity is
satisfieded and partial waves are indeed well unitarized by the IK method.
\section{Tracking the poles}
Poles in the complex ``s''-plane are provided by the complex zeroes of the determinant in the inverted term of Eq.~(\ref{eq_t0}). Since the determinant
is common to all entries of the unitarized matrix of amplitudes, poles should in principle be visible in all of them, but the coupling of a given channel
to a particular pole need not be the same, implying that they may be quite apparent in a given channel and difficult to see in another.\myspace
All dimensional quantities and plots are shown in units of the Planck mass. Thus, we define:
\begin{equation}
\bar{m} = \frac{m}{\masspl}, \quad \bar{\mu} = \frac{\mu}{\masspl}, \quad \bar{\Lambda} = \frac{\Lambda}{\masspl}, \quad \bar{s} = \frac{s}{\masspl^2},
\end{equation}
where $m$ is the mass of the scalaron, $\mu$ is the IR cut-off regulating the bad behavior of the partial wave in the extremes, $\Lambda$ is the
UV scale that has to be included in the unitarization process and $s$ is the Mandelstam variable that is promoted to a complex quantity.\myspace
We did not find significant variations  when changing the ultraviolet cut-off $\bar{\Lambda}$, From this point onwards, we fix $\bar{\Lambda} = 0.9$.
Then our analysis is restricted to two parameters: the infrared cut-off $\bar{\mu}$, and the scalaron mass $\bar{m}$, which is related to the
parameter $\alpha$ through Eq.~(\ref{eq_g_scalaronmass}).\myspace 
Numerically, one can reach extremely small values of the infrared cut-off, but this really makes no sense. 
Indeed, the appearance of infrared logarithms makes necessary to  place a lower bound on the IR scale and the $\mu\to 0$ limit cannot be taken in
a trivial manner.
As we discussed before, the natural expansion parameter of the theory is $\kappa^2/16\pi^2$ and higher momentum corrections will appear in the form 
$\kappa^2 s/16\pi^2$, but the iteration of IR divergences will presumably appear as
$(\kappa^2 s /16\pi^2)\log(s/\mu^2)$. In terms of the dimensionless variables previously introduced this is equivalent to
\begin{equation}
\frac{4}{\pi} \bar{s} \log(\frac{\bar{s}}{\bar{\mu}^2}).
\end{equation}
 We need this effective expansion parameter to be smaller than one, or at least not much bigger than one just to get qualitative results. If we place ourselves
in the sub-planckian region, simple considerations allow us to conclude that we cannot go to very small values of $\bar{\mu}$.  
 Specifically, the results presented in this section should be considered tentative as we explore values of $\bar{\mu}$
down to $10^{-20}$ and focus mostly on the observed trend.
\subsection{The scalaron}
\begin{figure}
\centering
\includegraphics[width=0.47\textwidth]{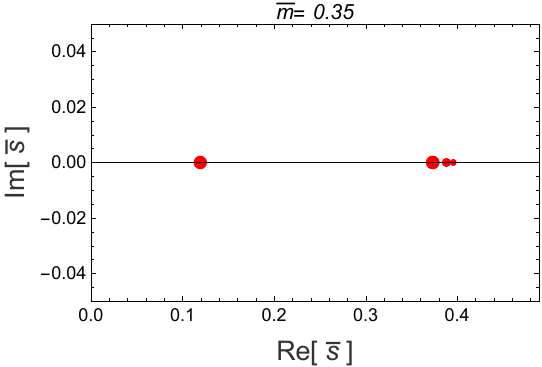}
\includegraphics[width=0.47\textwidth]{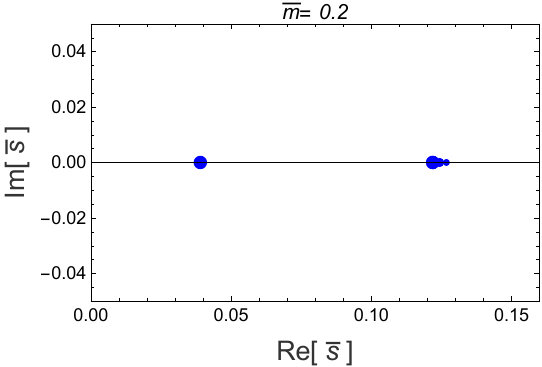} 
\includegraphics[width=0.47\textwidth]{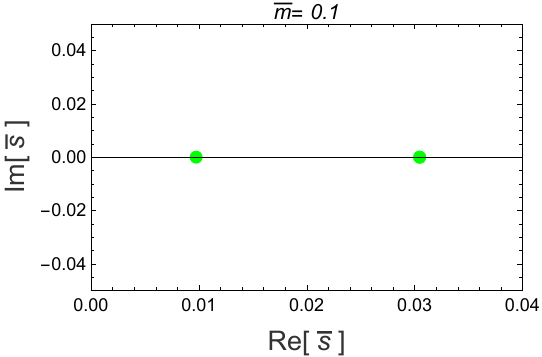}
\includegraphics[width=0.47\textwidth]{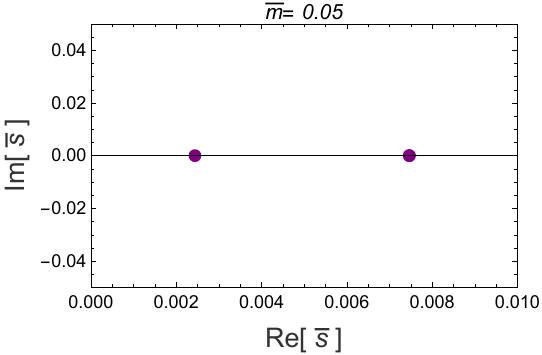} 
\caption{\small{Position of the scalaron and the $3m^2$ logarithmic branch point for various values of $\overline{m}$ and the indrared cut-off $\overline{\mu}$. In each panel, the values for the dimensionless IR regulator are $\overline{\mu}=10^{-10}, 10^{-15}, 10^{-20}$ and are identified with different pointsizes---bigger pointsizes represent bigger values of $\overline{\mu}$. In the last two panels, the position of the logarithmic branch point
    is unchanged by the regulator.}}
\label{fig_sc_3m2_real}
\end{figure}

\begin{figure}
\centering
\includegraphics[width=0.47\textwidth]{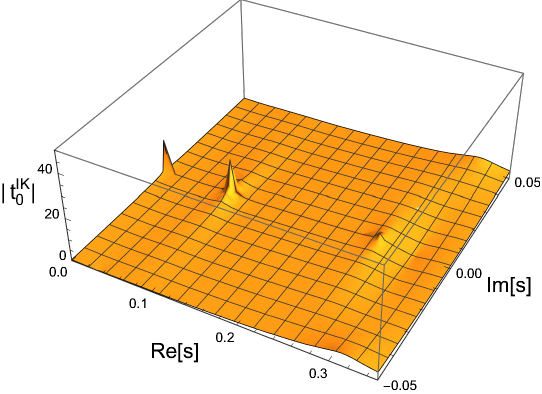}
\caption{\small{The scalaron pole and the logarithmic singularity at $3 \bar{m}^2$ in the second Riemann sheet for $\bar{m}=0.3$. The value of the IR regulator is $\bar{\mu}=10^{-5}$.}}
\label{fig_sc_3m2_complex}
\end{figure}

Our first task is to identify the pole due to the $J=0$ scalaron and also the branch point that is expected to appear in the $J=0$ partial wave due to
the one-scalaron exchange in the $t$ and $u$ channels. We do not expect to find the poles exactly at the location of the Lagrangian because in some
sense unitarization includes higher order effects and they can be displaced, but in any case not too far away.\myspace
A simple estimation of the $1\to 2$ one-loop diagrams involved shows that the expected width for the scalaron decaying into a pair of gravitons scales as $\sim m^7/M_{\text{pl}}^6$, yielding a negligible imaginary part in the complex $s$-plane for natural values of the scalaron mass. Indeed, after applying the unitarization method, we do not observe any displacement of the scalaron pole, neither along the real nor the imaginary directions in the complex $s$-plane.\myspace
Also, we naturally expect to visualize the poles just mentioned in the 22 channel; that is, the scattering of two scalarons yielding also two scalarons.
And indeed this is the case. Fig.~\ref{fig_sc_3m2_real} shows the position of the scalaron pole for a variety of Lagrangian values for $m$---recall that
we use dimensionless quantities. The plot also includes the variation due to choosing a range of values for the infrared cut-off. The conclusion is
obvious: there is almost no dependence on $\overline{\mu}$, especially for the pole of the scalaron at $\overline{m}^2$ due to $s$-channel exchange.\myspace
In Fig. \ref{fig_sc_3m2_complex} we also depict the second (adjacent) Riemann sheet to visualize the two mentioned poles for a particular value of $\overline{m}=0.3$.

\subsection{The graviball}
\begin{figure}
\centering
\includegraphics[width=0.49\textwidth]{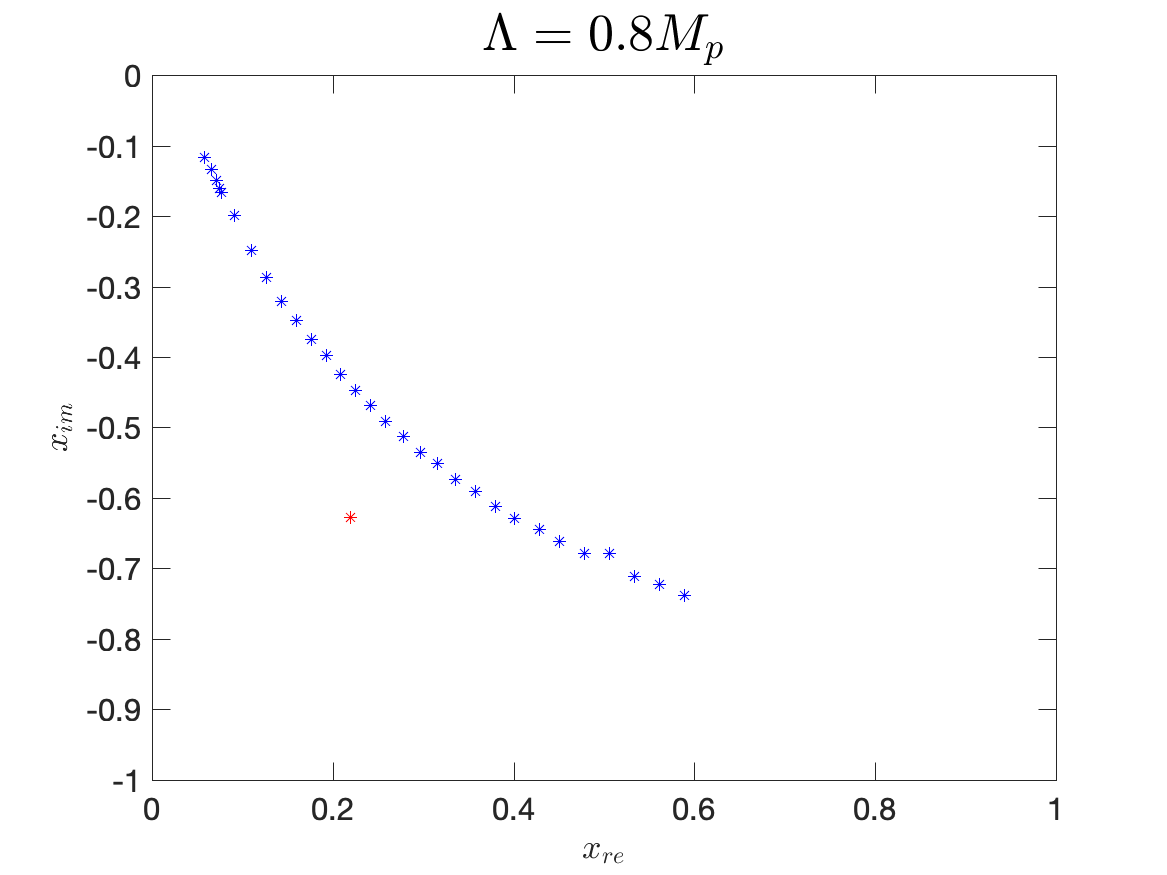}
\includegraphics[width=0.49\textwidth]{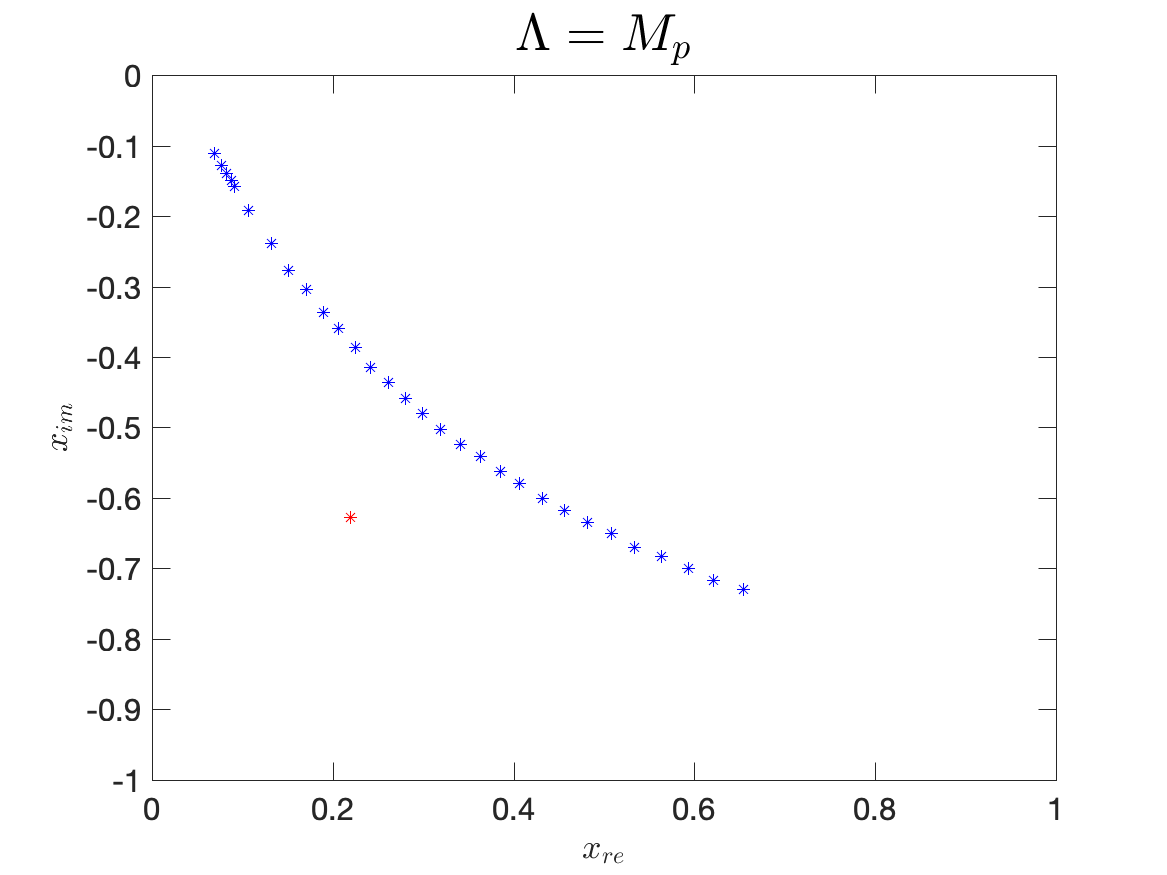} 
\caption{\small{Position of the graviball (blue points) for $\mu \in (0.001, 0.5)$ in pure gravity at leading order and two values of $\Lambda = 0.8 \masspl$ (left) and $\Lambda = \masspl$ (right). The results show a trend toward the origin, ruling out the possibility of it being a physical resonance. The pole from Ref.~\cite{Blas:2020dyg,Blas:2020och,Oller:2022ozd} is also indicated with a red point.}}
\label{fig_g_graviball_sc}
\end{figure}

\begin{figure}
\centering
\includegraphics[width=0.47\textwidth]{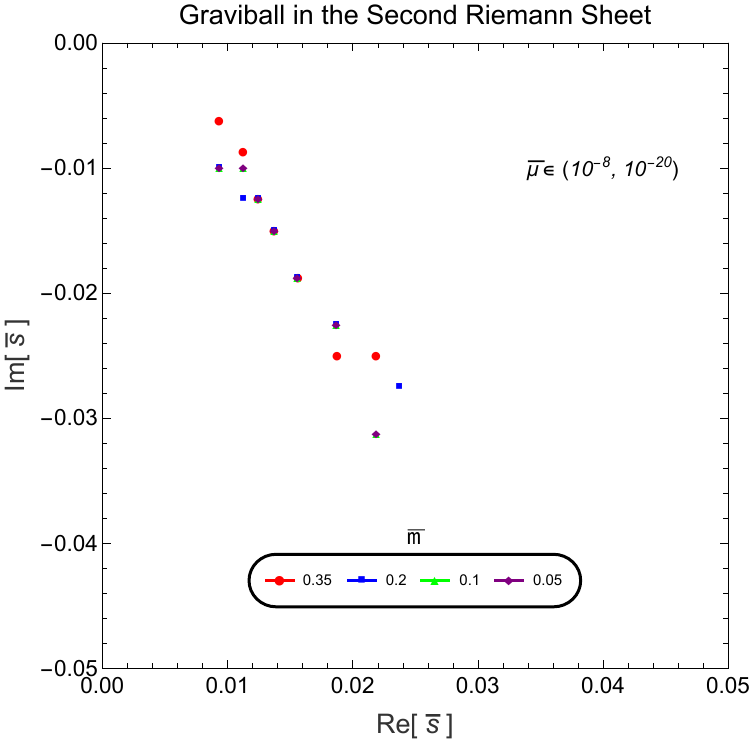}
\includegraphics[width=0.47\textwidth]{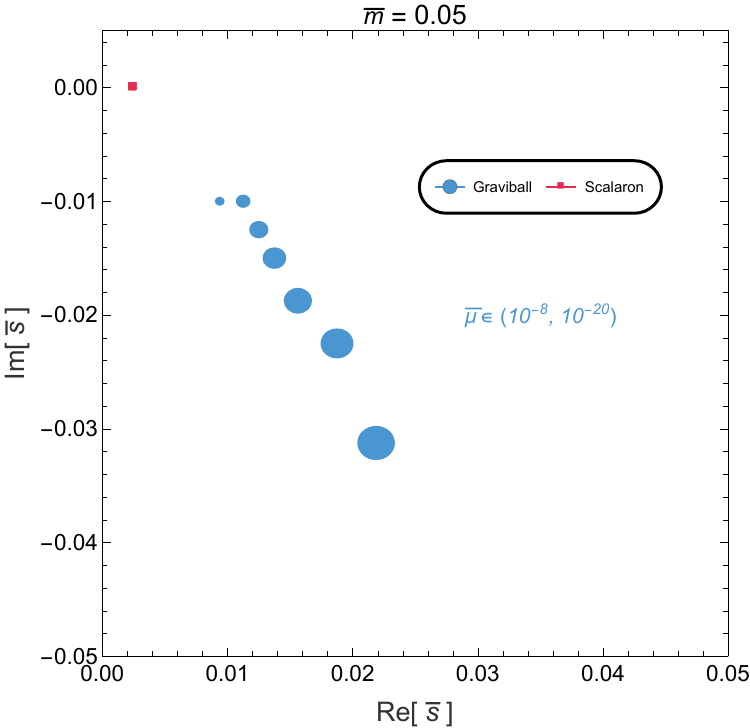} 
\caption{\small{A comparison between the displacementof the graviball pole under changes in the infrared regulator for different scalaron masses (left) and for $\overline{m}=0.05$ (right). In the left panel, the values of $\overline{\mu}$ vary from $10^{-8}$ to $10^{-20}$, the latter producing graviball locations closer to the origin for all values of the scalaron mass. The right panel show the same for a specific value of $\overline{m}$ and we show in a red squar the position of the scalaron too, independent of $\mu$. }}
\label{fig_g_graviball_cc}
\end{figure}

The conspicuous pole in the second Riemann sheet corresponding to two graviton scattering, found in
Ref.~\cite{Blas:2020dyg,Blas:2020och,Oller:2022ozd} and confirmed in Ref.~\cite{Delgado:2022qnh,Delgado:2022uzu},
should also be  seen in the coupled channel calculation, where it could be expected to mix somehow with the scalaron. However, we have already
seen that the latter is fairly insensitive to the infrared cut-off.
The graviball appears at very low energies in the $J=0$ partial wave of graviton-graviton scattering in pure gravity. It has been  interpreted as a
putative 2-graviton compound by gravitational forces, but its nature as a true dynamical resonance was challenged because is very far from the
real $s$ axis and thus not meeting the usual criteria for genuine resonances, and, at least in pure gravity, being highly dependent on the IR
regulator. Nevertheless, some phenomenological speculations have already been proposed (see for instance Ref.~\cite{Guiot:2020pku}).\myspace
The authors of Refs.~\cite{Blas:2020dyg,Blas:2020och,Oller:2022ozd,Oller:2024neq} argue in favour of the presence of a resonance, whose position
depends on a parameter $a $ (their notation) that regulates infrared divergences appearing for the propagation of massless modes. This parameter redefines
the energy scale $Q^2 = \pi G^{-1}/\log a $, which is characteristic of the elastic graviton process. The authors find the position of the
graviball at $s = \left(0.22 - i 0.63\right)(G \log a)^{-1} $. As already said, this could not be classified as a genuine
resonance because it does not satisfy a natural relationship between its mass and width \textit{à la Wigner}.\myspace
Instead, the authors of Refs.~\cite{Delgado:2022qnh,Delgado:2022uzu} also identify this structure but describe it as an artifact due to the introduction of
the parameter $\mu$, which regulates partial waves in the infrared, analogous to $a$ in the previous case. Using their unitarized analysis with the
IAM at next-to-leading order, they determine that the position of this pole is consistent with $s = 0$ as $\mu\to 0$, distancing it from any physical
significance as a dynamical state. Of course this statement requires  qualification because infrared divergences are definitely present in
graviton scattering and the limit $\mu\to 0$ cannot be simply taken as higher order contributions would become more and more relevant. One has
to stay with values of the infrared cut-off that do not conflict with perturbation theory. The authors in Ref.~\cite{Delgado:2022qnh,Delgado:2022uzu}
did two things: first they found a value of their IR regulator that grossly coincided with the pole found by the authors in
Refs.~\cite{Blas:2020dyg,Blas:2020och,Oller:2022ozd,Oller:2024neq}; second, they reduced the value of the infrared cut-off and convinced theselves that the ``pole'' was
moving fast towards the origin and therefore coould be considered an artifact of the IR regulation procedure.\myspace
It seems pertinent to recall that self interactions of the graviton involve derivatives (see Eqs. (\ref{eq_g_Sfull}) and (\ref{eq_g_S3_and_S4})) impying that the theory is a free one in the infrared and the only degrees of freedom that can appear at $s=0$ are the ones already existing in the Lagrangian. Consequently, the fact that a putative dynamical resonance is driven to $s=0$ as the IR cut-off is removed signals that this resonance cannot possibly exist because the theory is a free one in that limit.\myspace
Our results align here with this latter scenario. First, we verified that, indeed, using a single channel for unitarization
(but now using the IK-matrix method); that is, neglecting the mixing with the scalaron in the $J=0$ channel, the main characteristics of the graviball found
in Refs.~\cite{Delgado:2022qnh,Delgado:2022uzu} are reproduced. That is, as the infrared cut-off tends to zero, $\mu \to 0$, the pole moves closer
and closer to the origin. Secondly, that this behaviour persists in the coupled channel formalism when the coupling to a genuine scalar degree of
freedom is considered.\myspace 
In Fig.~\ref{fig_g_graviball_sc}, the position we obtain for the graviball (blue points) in the single-channel approximation is shown for values
of $\bar{\mu} \in (0.001, 0.5)$ and two UV cut-offs, $\bar{\Lambda} = 0.8$ and $\bar{\Lambda}=1.0$. Recall that quantities with a bar are in
units of the Planck mass and are therefore dimensionless. We also show, for comparison, the position of the pole from Ref.~\cite{Blas:2020dyg,Blas:2020och,Oller:2022ozd}.
\myspace
Now we repeat the exercise in the coupled-channel formalism. As discussed before in the text, the functions with coupled channels depend not only on
the IR regulator, but also de mass of the scalaron. The results for the position of the graviball are shown in Fig.~\ref{fig_g_graviball_cc}.
The left panel displays the pole location for various values of IR regulator and the scalaron mass, with the latter ranging from $\overline{m}=0.35$
down to $\overline{m}=0.05$. The IR regulator values are $10^{-8},\, 10^{-10},\cdots, 10^{-20}$ and, for each series, the smaller the IR regulator,
the closer the pole moves toward the origin. This behavior is shown explicitely in the right panel, which focuses on $\overline{m}=0.05$ (the closest
value in our scan to that of the Starobinsky model). Here, the graviball pole positions are plotted with point size encoding the value of the
IR regulator---larger points correspond to larger regulators.\myspace
In Fig.~\ref{fig_g_graviball_cc} we also show the scalaron pole, that is located at its expected position $\overline{s}=\overline{m}^2=0.0025$ no matter
the choice of the IR regulator. This different behavior for the graviball and the scalaron indicates the difference between a proper resonance,
as it is the case of the scalaron, and what we consider an artifact. 
\subsection{Other resonances}
\begin{figure}
\centering
\includegraphics[width=0.47\textwidth]{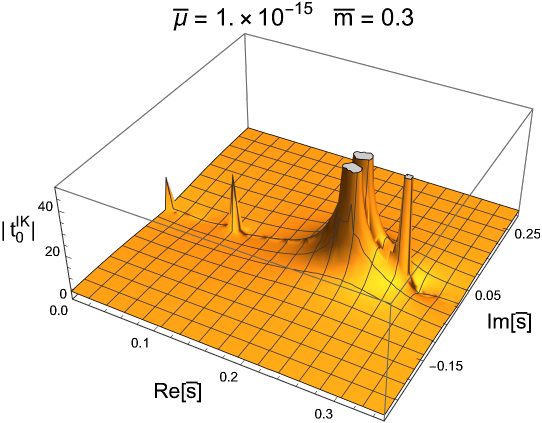}
\includegraphics[width=0.47\textwidth]{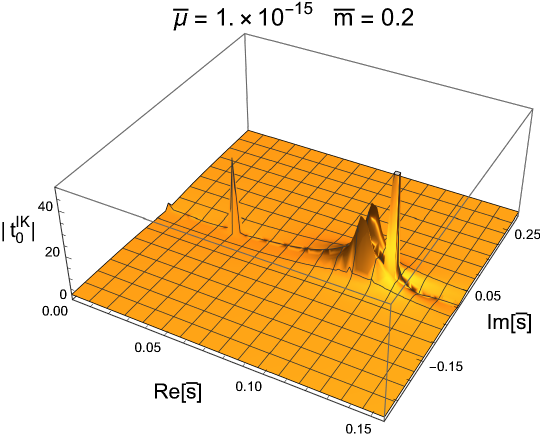}
\caption{\small{Unitarized 22 partial wave for $\bar{\mu}=10^{-15}$ and for different values of the dimensionless scalaron mass $\bar{m}=0.3$ (left) and $\bar{m}=0.2$ (right). The region of the complex plane above the real axis corresponds to the physical Riemman Sheet (1RS) and the region below corresponds to the unphysical Riemann Sheet (2RS), where the amplitude has been continued analitically.}}
\label{fig_twins_samemu}
\end{figure}

\begin{figure}
\centering
\includegraphics[width=0.47\textwidth]{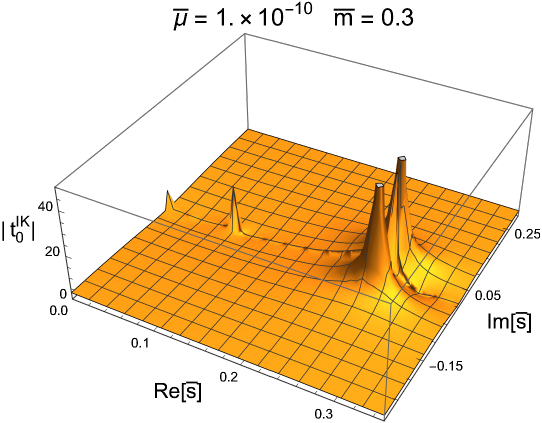}
\includegraphics[width=0.47\textwidth]{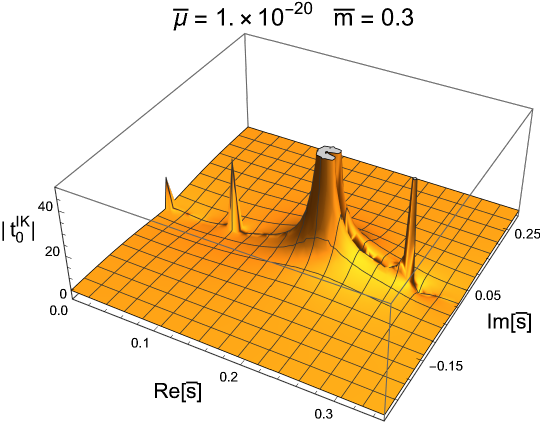}
\caption{\small{Unitarized 22 partial wave for $\bar{m}=0.3$ and for different values of the dimensionless IR regulator $\bar{\mu}=10^{-10}$ (left) and $\bar{\mu}=10^{-20}$ (right). The region of the complex plane above the real axis corresponds to the physical Riemann Sheet (1RS) and the region below corresponds to the unphysical Riemann Sheet (2RS), where the amplitude has been continued analitically.}}
\label{fig_twins_samemass}
\end{figure}

Quite interestengly, the study of the partial waves in the second Riemann sheet reveals further structures. This is best seen in the 22 partial wave after unitarization.
In fact, without making use of the coupled channel formalism, the 11 amplitude coincides with the one derived of the pure Einstein-Hilbert theory,
where no proper dynamical singularity seems to be present (only the graviball, which progressively vanishes as the IR cutoff tend to zero).
Therefore additional singularities may appear in the 11  partial wave
only after mixing with the 22 one and they tend to be fairly hard to see, indicating a small overlap of these resonances that are visible in the 22 channel with
the two graviton initial state.\myspace
Examining the 22 partial wave in the second Riemann sheet, we detect a singularity, clearly different from the ones due to the scalaron or to the
scalaron exchange (that always lie on the real $s$ axis) and the graviball. This new singularity moves closer to the real axis as $\bar\mu$
decreases and it moves closer to the imaginary $s$ axis as $\bar{m}$
decreases; its real part is always located above the scalaron pole and the threshold at $s=4 \bar{m}^2$. Perhaps surpisingly, it seems to be associated to
a mirror singularity in the first (physical) Riemann sheet. This is clearly visulized in Fig.~\ref{fig_twins_samemu} for various values of the scalaron mass $\bar{m}$.
\myspace
However, there is actually nothing really  odd here. If we consider the 22 partial wave by itself; that is, decoupling it from the 11 channel, it possesses a
branch cut starting
at $4m^2$. It turns out that the real part of the singularity being discussed now is always below threshold and therefore, the singularity for negative
and positive imaginary parts are in fact mirror resonances in the same Riemann sheet, something that is fine and relatively frequent. In the energy complex plane
(as opposed to the $s$ complex plane) the mirror image for positive values of the imaginary part of $s$ actually lies quite far away from the physical region.\myspace
Consequently, everything points in the direction  that we are dealing with a genuine dynamical resonance. Also interestengly, and unlike the so-called graviball,
it does not approach the origin when the infrared cutoff is progresively removed, but it becomes narrower and closer tho the positive real $s$ axis.
As said it always stays for values of the real part of $s$ below the branch point at $4 \bar{m}^2$. This makes this resonance somewhat unusual as one is
more familiar with the situation where resonances are above threshold (in which case the mirror resonances would not be on the same physically relevant
Riemann sheets). This state is heavier than the scalaron so it would be possible to interpret is as an excited state, needed to unitarize the amplitudes.\myspace
In fact, given the way it approaches the positive real axis, everything suggests that the would be resonance could be a bound state of
two scalarons. In the limit when the IR cutoff
is removed, it would be stable and it could only decay to two gravitons after the coupling to the 11 channel is reinstated. The evolution of this state for
a given scalaron mass as a function of the infrared cut-off is shown in Fig.~\ref{fig_twins_samemass}.
\myspace
Of course, when the coupled channel formalism is implemented, the analytic structure becomes more involved as the cuts from the various processes are superposed.
However, for small values of the infrared cut-off, there is little coupling between the 11 and 22 channels because the channel 
12 is much  smaller that 11 or 22. In practice, the dynamical resonance just found stays there, more or less in the same place in the 22 channel, and it 
has no visible impact on the 11 one.\myspace
It goes without saying that one should take the previus results {\em cum grano salis} because as we have repeatedly stated the cut-off cannot be really removed and
it cannot be made too small, lest the perturbative series explode. But the tendency is very clear.
\section{Unitarized amplitudes}
\begin{figure}
\centering
\includegraphics[width=0.65\textwidth]{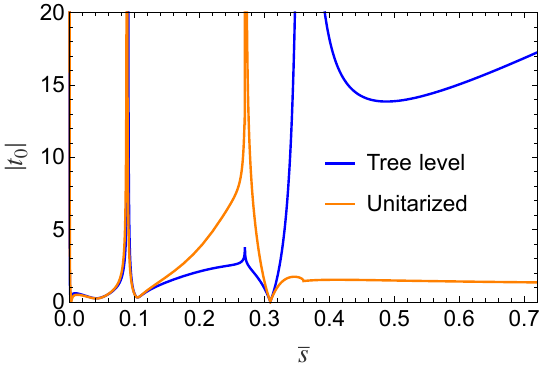}
\caption{\small{Absolute value of the physical tree level (blue) and unitarized (orange) partial wave in the 22 channel and in the $J=0$ projection with the Mandelstam variable $s$ in units of the Planck mass squared, $\overline{s}$. The values for the dimensionless IR regulator and scalaron mass are $\overline{\mu}=10^{-10}$ and $\overline{m}=0.3$. The scalaron pole at $\overline{s}=0.09$ and the logarithmic singularity at $\overline{s}=0.27$ are evident for both partial waves. At 
$\overline{s}=0.36$ the two scalaron channel opens.}}
\label{fig_unit_vs_nounit}
\end{figure}

\begin{figure}
\centering
\includegraphics[width=0.95\textwidth]{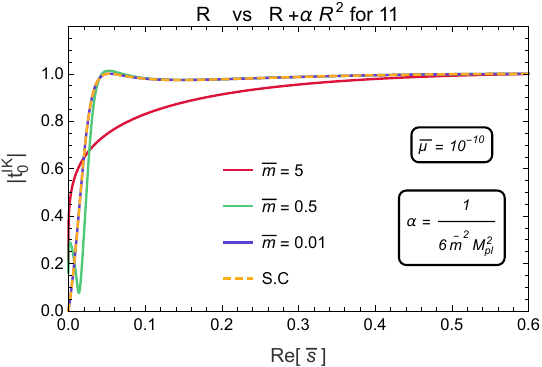}
\caption{\small{Absolute value of the unitarized amplitude for  $\left(R\right)$ and  $\left(R+\alpha R^2\right)$ for the 11 channel. The result for $R$ coincides with the 11 amplitude in the single-channel approximation and it is 
shown as a dashed orange line, while the next-to-leading-order result, requiring the coupled-channel formalism, is shown as solid lines for different values of the scalaron mass.}} 
\label{fig_RvsR2_11}
\end{figure}

\begin{figure}
\centering
\includegraphics[width=0.95\textwidth]{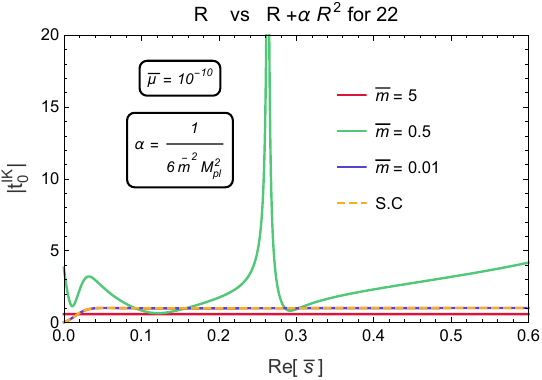}
\caption{\small{Absolute value of the unitarized amplitude  for the  22 channel (right). The leading-order result, obtained in the single-channel approximation, is shown as a dashed orange line, while the next-to-leading-order result, requiring the coupled-channel formalism, is shown as solid lines for different values of the scalaron mass.}} 
\label{fig_RvsR2_22}
\end{figure}

We insert a very small positive imaginary part in $\overline{s}$ in order to be on the physical Riemann sheet and visualize the partial waves for
physical values of its argument  In the absence of dynamically-generated scalar resonances close to the real axis,
the only structures that we expect to find in the physical region 
are the poles inherent to the low-energy theory, i.e. before unitarization, and discontinuities at threshold. This is true for the unitarized partial
wave due to the scalar matrix $t_0^{(2)}$ 
multiplying on the right in Eq.~(\ref{eq_t0}).\myspace
This is precisely the situation seen in Fig.~\ref{fig_unit_vs_nounit}, which displays both the non-unitarized (blue) and unitarized (orange) scalar waves in the 22 channel for the 
values $\overline{m}=0.3$ and $\overline{\mu}=10^{-10}$. At $\overline{s}=\overline{m}^2=0.09$, the pole corresponding to the scalaron is visible in both cases. 
The logarithmic singularity at $\overline{s}=3\overline{m}^2=0.27$---arising from the scalaron exchange in the $t$ and $u$-channels---is also shown for both curves. 
The production threshold of the scalaron pair, located at $\overline{s}=4\overline{m}^2=0.36$, is also noticelable in both curves: as a singularity in the tree-level scalar wave, and as a 
glitch that prevents  non-unitarity behavior in the physical region for the unitarized case. All these features appear at their corresponding locations for other values of the scalaron mass and the IR regulator.
Similar features are observed for other values of the parameters.\myspace
As we can see ---and was expected--- the unitarization process tames the high enery behaviour of the 22 amplitude, making it tend to a constant value. The same behaviour is observed in the
11 amplitude, that corresponds to the process $hhhh$, i.e. the elastic scattering of two gravitons. Incidentally, its asymptotic value coincides with one one of pure gravity (after unitarization).
\myspace
We want to draw a comparison between the leading-order and next-to-leading-order calculations in pure gravity. As explained earlier in the manuscript, performing 
the conformal transformation in Eq.~(\ref{eq_g_conformaltrans}) and adding the scalaron is equivalent to computing the next-to-leading-order term $\alpha R^2$, with $\alpha$ defined in Eq.~(\ref{eq_g_scalaronmass}). 
The task now is therefore to compare the unitarized calculation for $hhhh$ at tree level, Eq.~(\ref{eq_t0_sc}), with this including coupled channels, Eq.~(\ref{eq_t0}), for different values of the 
scalaron mass. The results are shown in Figs.~\ref{fig_RvsR2_11} and \ref{fig_RvsR2_22} for both 11 and 22 channel, respectively.
\myspace
Interestingly,  for the lower values of the scalaron mass, a decoupling of the two channels occurs: the crossed channel vanishes. Also, channels 11 and 22 coincide (this can be checked analytically too even before unitarization) 
and their value tends to coincide with the one of pure $R$ gravity. 
At first sight, this may appear contradictory since small values of the scalaron mass correspond to large values of the parameter $\alpha$ that controls the relevance of the $R^2$ term. In fact in the limiting case only the $R^2$ 
term would matter.  Also, the equality between the two channels holds for large values of $s$, as in practice this coincides with the small $m$ limit. All these results hold independently of the value of
the IR cut-off provided that is very clearly separated from the physical region.\myspace
For larger values of the scalaron mass, commensurate with the Planck mass, the $hhhh$ amplitude shows an enhancement  at low values of $s$. As the mass of the scalaron grows, the subplanckian part of the amplitude
becomes progressively featureless. Yet it tends to the same universal asymptotic value. Moving the scalaron mass to the superplanckian region smoothens  the amplitude further. However, it does not tend to the case where one 
sets $\alpha=0$ directly in the lagrangian. We interpret this as a manifestation of non-decoupling of the scalar mode that makes both limits non-commuting.\myspace
\section{Conclusions}
We have presented unitarized partial waves for the $R+\alpha R^2$ gravity theory, a.k.a. the Starobinsky model, a particular class of $f(R)$  theories. 
Perturbatively, for generic values of the coupling $\alpha$ this model presents three states:  two helicities for the graviton (which may combine to $J=0$ in graviton-graviton scattering) and a genuine spin zero state, the scalaron.
Projection of the partial waves in the scalar channel leads to a rather rich structure that has been studied in the framework of the (coupled channels) improved K matrix unitarization mechanism. 
The analysis reveals the expected zero-width poles corresponding to the scalaron and
the (logarithmic) pole due to the $t$ and $u$ scalaron exchange that are most visible in the $\varphi\varphi\varphi\varphi$ amplitude and have very small overlap with the two graviton scattering $hhhh$. 
The latter is as a whole not very affected by the coupled channel and shows characteristis somehow similar to those in pure Einstein gravity (i.e. $\alpha=0$), but some differences 
have been seen. Incidentally, the limit $m\to\infty$ ($\alpha\to 0$) seems to be discontinuous
because the scalaron shows signs of non-decoupling, something that can be understood from the couplings in the scalar potential. The limit $m\to 0$ ($\alpha\to\infty$) is interesting too and quite subtle. See Ref.~\cite{Casado-Turrion:2024esi} and references therein for a recent discussion about the pure $\alpha R^2$ theory on
a Minkowski background.\myspace
One of purposes of this study was to identify the dependence 
of the various complex structures appearing in the second Riemann  sheet on the infrared cut-off.. 
The IR singularities are regulated by means of a physical cut-off in the angular distribution, a procedure that was verified to be consistent with perturbative unitarity. Indeed, as was somehow 
expected, a genuine physical resonance such as the scalaron shows almost no dependence on the infrared cut-off. On the contrary, both partial waves after unitarization displays a pole, very far from the real
axis, that is very strongly dependent on the IR cut-off and tends to move quickly to $s=0$ as the cut-off tends to zero.\myspace
We have analyzed the range of values for the IR cut-off that are, at least nominally, compatible with perturbation theory and adhered to those. The study has some other limitations: it is difficult
to go to very low values of the mass for the scalaron  because the structure in the second Riemann sheet become rather hard to see, but that limit tends clearly, at least at the level 
of the amplitudes,  to pure Einstein-Hilbert gravity  The range of values of the Starobinsky model is about as low as one can get numerically. It is also delicate to go to large masses for  the
scalaron mass, something that can be understood in terms of non decoupling. \myspace    
An additional pole is clearly seen in the $\varphi\varphi\varphi\varphi$ (or 22) amplitude below threshold. Actually an inocuous mirror resonance is seen on the physical Riemann sheet too. The dynamical resonance on the second Riemann sheet seems genuine and
physical and should be considered as a valid prediction of the $R + \alpha R^2$ model after unitarization. In fact, all evidence points to the fact that it corresponds
to a bound state of scalarons. Of course further work with other unitarization methods would be very convenient in order to check
the stability of the prediction.\myspace
As for the amplitudes, they all tend to the same asymptotic value for scalaron masses in the subplanckian region. This limit coincides
with the one of Einstein-Hilbert, after unitarization, and show a 
limiting constant behaviour independent of $s$. \myspace

\section*{Acknowledgements}
 The financial support from
the State Agency for Research of the Spanish Ministry of Science, Innovation and Universities through the “Unit of Excellence Maria de Maeztu 2025-2028”
award to the Institute of Cosmos Sciences (CEX2024-001451-M) and from projects PID2022-136224NB-C21 and PID2022-137003NB-I00 is acknowledged. We also acknowledge
the suppport of the Catalan Government through grant 2021-SGR-249. 
A.D. thanks J. R. Peláez for very useful discussions.
D.E. thanks the hospitality of the theory group at Laboratorio Nazionale di Frascati, where this work came to its final form.

\begin{small}

\bibliographystyle{utphys.bst}
\bibliography{bibliography}

\end{small}

\end{document}